# Assessment of OpenStreetMap Data - A Review

Sukhjit Singh Sehra
Assistant Professor
GNDEC, Ludhiana, India

Jaiteg Singh
Associate Professor
CIET, Punjab, India

Hardeep Singh Rai
Professor
GNDEC, Ludhiana, India

**ABSTRACT**

The meaning and purposes of web has been changing and evolving day by day. Web 2.0 encouraged more contribution by the end users. This movement provided revolutionary methods of sharing and computing data by crowdsourcing such as OpenStreetmap, also called "the wikification of maps" by some researchers. When crowdsourcing collects huge data with help of general public with varying level of mapping experience, the focus of researcher should be on analysing the data rather than collecting it. Researchers have assessed the quality of OpenStreetMap data by comparing it with proprietary data or data of governmental map agencies. This study reviews the research work for assessment of OpenStreetMap Data and also discusses about the future directions.

## 1. INTRODUCTION

The Web started with web 1.0 or "read only" web, but Tim O'Reilly [1] discussed the concept of Web 2.0 or "read-write" web. The client side technologies such as Ajax and JavaScript framework are used for Web 2.0 development [1, 2]. Web 2.0 encourages greater collaboration among internet users and other users, content providers, and enterprises [3]. This movement provided revolutionary new methods of sharing and computing data by crowdsourcing movement similar to wikipedia [4, 5, 6]. In regard to the geographical data the crowd-sourced movement is known as VGI (volunteered geographic information) [8], others call it collaborative mapping [9], so it is a special case of this web phenomenon and has been applied in many popular websites such as Wikimapia, OpenStreetMap(OSM), GoogleMap, Flickr [10]. The two major data providers are NavTeq and TeleAtlas. However, these data are costly, quickly outdated and restricted to specific areas covered by the data acquiring companies. Large companies have invested large sums of money to purchase smaller companies to acquire their data e.g. in 2007, Nokia acquired NavTeq, in 2006, Microsoft acquired the Imagery and Remote Sensing Company Vexcel [12].

The website of Openstreetmap is a collaborative project to create a free editable map of the world, using crowdsourced approach. Two major driving forces behind the establishment and growth of OpenStreetMap have been restrictions on use or availability of map information across much of the world and the advent of inexpensive portable satellite navigation devices. GPS enabled smartphones based on iOS (72 millon Iphones worldwide) or Android (One billion users worldwide) [15] made easy for users to contribute to crowd sourced openstreetmap. OpenStreetMap is based on the concept of crowdsourcing, also called wikification of GIS [13], which encourages the volunteers worldwide to contribute through the collection of geographic data. The data of OpenStreetMap is useful because firstly, the data is completely free with an open content licence. Secondly, it is constantly being updated by the subscribed users who can also add points of interest important to them. OpenStreetMap has been used during earthquake situation in Haiti named as Haiti Crisis Map. Finally, OpenStreetMap has the potential to establish volunteers from all over world including less developed regions, where obtaining data can be difficult for most commercial mapping companies [14]. Till July, 2013 nearly 1,300,000 have registered as users/contributors to OpenStreetMap [11], fig 1 show the graph. To compete with OpenStreetMap, Google introduced a tool called Map Maker in 2008, that enabled users to contribute data themselves. This tool was only available for areas with no or little commercial data coverage, e.g. India, Pakistan, Iceland and within a short time, large areas were mapped in this crowdsourcing manner [12].

Next section discusses about the assessment work on Openstreetmap. Nearly every researchers have compared the OpenStreetMap Data with actual ground data and some compared it with the proprietary data sets. Last section is conclusion and future directions.

## 2. QUALITY ASSESSMENT PARAMETERS

For OpenStreetMap to replace proprietary maps, the quality of map data has to assessed, so many researchers started comparing OSM data with ground data available with different governmental agencies. The Geographical information from different sources may be compared based following parameters [10, 14, 18]:-

(1) ineage: - It describes the record of the spatial data, which includes the detailed description of the source material from which the data were derived, the methods of derivation, the dates of the source material, and all transformations involved in producing the final digital files.

[(2)] ositional Accuracy: - Positional accuracy defines the closeness of locational information to the true position i.e. the absolute and relative accuracy of the positions of geographic features. The Absolute accuracy is the closeness of the coordinate values in a dataset to values accepted as or being true and the Relative accuracy is the closeness of the relative positions of features to their respective relative positions accepted as or being true.

[(3)] ompleteness: - It describes the degree to which geographic features, their attributes and their relationships are included or not included in spatial dataset. In addition it also include information on selection criteria, definitions used, and other relevant mapping rules.

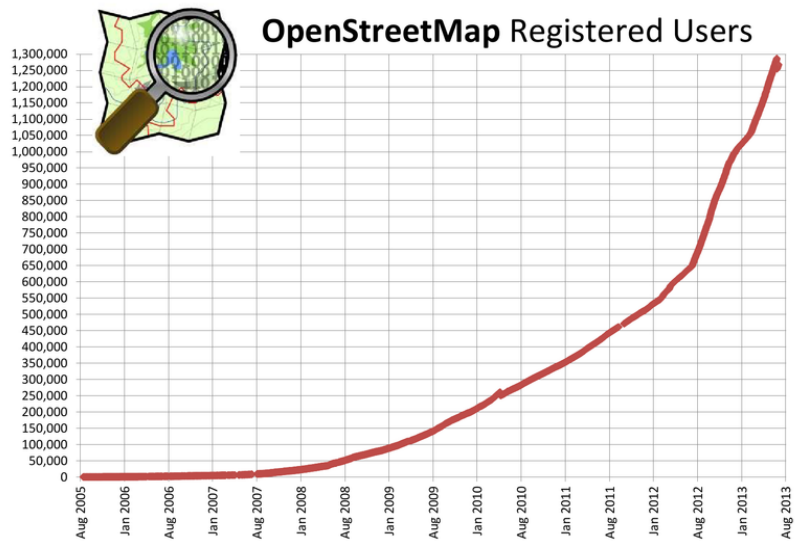

**Fig. 1. Registered Users**

Some researcher also analysed the spatial data quality using parameters in addition to discussed above, which are Thematic Accuracy, Temporal Accuracy, Logical consistency, Semantic accuracy, Usage/purpose/constraints, Variation in quality, Metaquality, and Resolution. Often one dataset may be superior to other datasets in one, but not all aspects.

## 3. ASSESSMENT WORK

OpenStreetMap is based on the concept of crowdsourcing and founded by Steve Coast from UK in 2004. OpenStreetMap produces huge spatial data, with less effort. and researchers are working on the devising method to use the data rather than collecting the data. OpenStreetMap produces labelled data, When labeled data is easy to come by, the focus of the researcher would be on working with the labelled data rather than collecting it. This impacts not only which problems researchers choose to work on, but also the learning methodology they use to approach them.
In this sense, crowdsourcing provides a two-part cost savings: greater ability to realize traditionally-claimed savings of active learning, as well as reduced cost of crowd annotation vs. traditional annotators. A second important benefit will again be the implications for use of labeled vs. unlabeled data for training when labeled data is plentiful. Instead of comparing to past supervised learning curves, researchers may instead consider past learning curves for active learning, which will be steeper in comparison. In many prior studies, it has been concluded that there is greater potential for active learning than supervised learning to benefit from crowdsourcing [16, 17].
Haklay, in 2008, [18] analysed OSM data compared Great Britain, and ordnance survey (OS) geodata with OSM data. In 2009, Ather [14] extended this work to the OS Master Map for selected parts of London. He additionally compared completeness of road names. Then, in 2010, Haklay [19] in his research work buffered British Ordnance Survey data to determine what percentage of the OSM roads were covered. A commonly applied technique for matching different road networks is graph matching [20]. The analysis of Germany started with comparison of commercial mulinet proprietary map data from TomTom [21, 22] compared with street map data from different proprietary geodata providers.

In 2011, Ciepluch [30] discussed that not everyone contributes data of the same quality, due to lack of practice and knowledge which can be improved by practice and experience in map making. Ludwig [22] described a methodology to compare OSM street data with Navteq for all populated roads in Germany. The methodology was based on a matching between the street objects of OSM and Navteq adopted from [23] which allows for object-wise comparison of geometries and thematic attributes. Finally, they calculated relative quality measures: relative object completeness, relative attribute completeness, difference in speed limits and positional differences. Another researcher [24] statistically analysed the routing process using OpenStreetMap road data of the inner city of Hamburg. A similar approach was used in France to analyse OSM data [25]. The results of this research showed the advantage and flexibility, but also concluded the problem of the heterogeneity of the data specifically for France. In same year, the first study that analysed the quality of OSM outside of Europe were conducted [26]. In this research work the OSM project data had been compared with proprietary data from TomTom (TeleAtlas) and Navteq for the entire state of Florida (USA) and four specific cities within the USA. In comparison to the results for Germany or England, the discrepancies between the rural and urban areas in the USA showed an opposite tendency. In Florida, the rural data was, in parts, even more complete than that of the proprietary datasets in the relative comparison conducted. Zielstra [27] compared the amount of pedestrian-related data between freely available sources, i.e., OSM and/or TIGER, and proprietary providers, i.e., Tele Atlas Multinet and/or Navteq Discover Cities, and analysed its effect on modeling ?spatial aspects of transit accessibility" for pedestrian in five US and four German cities. They concluded that integration of pedestrian-only segments can lead to a more realistic assessment of service areas when compared to using networks that contain only streets that are passable by cars and that the assessment of VGI data quality, especially OSM data, is an ongoing issue of high importance for successful geo-applications [7, 19].
Other analyst in 2012 [28] assessed the effect of network data integration from different sources on the length of computed shortest paths for pedestrians. Their results showed that the combined use of network datasets significantly reduces shortest path distances compared to the use of single datasets. They concluded that data integration leads to an increased value for users of pedestrian routing applications but that combining OSM and other commercial datasets cannot be considered for implementation due to current licensing issues. Neis [29] assessed the completeness of the OSM street network via a relative comparison (street network length, no. streets without names, no. turn restrictions) between OSM and a commercial dataset provider (TomTom formerly known as Tele Atlas). They noted though that for comparison the TomTom dataset is suitable only for street network data for car-specific navigation. They also evaluated logical consistency using an internal test, whereby topological and thematic consistency is determined. Concerning turn restrictions (176,000 in TomTom; 21,000 in OSM in Germany in June 2011) they discussed that although the number of turn restrictions available in the OSM dataset is continually increasing, it will probably take several more years before OSM achieves the same level as TomTom, based on the current status and development?. Apart from England, no studies have been conducted to date over a period of several years and for an entire country [19].

In 2013 many researchers have been aggressively working in the area of assessment of OpenStreetMap, but research work for India has not initiated yet.

## 4. CONCLUSIONS & FUTURE WORK

This review paper concludes that OpenstreetMap is generating huge dataset with the help of non-commercialised users, with varying level of experience. So the assessment becomes vital to give maturity to OpenStreetMap data. The results from the finding show that because of varying level of user experience the data is not error free, and also mapped areas depend upon the contribution by the users. But general trend is that number of absolute and relative errors is falling.

The discussed approaches in this paper are offline approaches of checking the data and correcting the OSM data afterwards. Another method which still needs attention from the researchers community is online quality check or anomaly detection engine, that can check for the quality of the map while it is being uploaded by the user. The idea may be to create automated model that can spot mistakes. In order to do this some machine learning techniques with several user parameters being: the age of the user, the number of edits, what happens in a changeset etc can be taken into consideration as discussed by Neis [29]. Further some of the researchers have compared data proprietary data sets and given their results. The ground reality of those proprietary data sets may also be checked.